\documentclass[12pt,a4paper]{article}

\usepackage[top=2.5cm,bottom=2.5cm,left=2.8cm,right=2.8cm]{geometry}
\usepackage{amsmath,amssymb}
\usepackage{graphicx}
\usepackage{float}
\usepackage{caption}
\usepackage{subcaption}
\usepackage{booktabs}
\usepackage{setspace}
\usepackage{hyperref}
\usepackage{authblk}
\usepackage{xcolor}
\usepackage{bm}
\usepackage{array}
\usepackage{tabularx}

\hypersetup{
    colorlinks=true,
    linkcolor=blue!60!black,
    citecolor=blue!60!black,
    urlcolor=blue!60!black
}

\newcommand{\nue}{\nu_e}
\newcommand{\numu}{\nu_\mu}
\newcommand{\nutau}{\nu_\tau}
\newcommand{\dcp}{\delta_{\rm CP}}
\newcommand{\Eg}{E_{\rm global}}
\newcommand{\DS}{\Delta S}
\newcommand{\DSp}{\Delta S_{+}}
\newcommand{\DSm}{\Delta S_{-}}
\newcommand{\DSn}{\Delta S_{\nu}}
\newcommand{\DSnb}{\Delta S_{\bar\nu}}
\newcommand{\NO}{{\rm NO}}
\newcommand{\IO}{{\rm IO}}
\newcommand{\dm}[1]{\Delta m^2_{#1}}

\title{\textbf{Tripartite Entanglement as a Probe of Neutrino Mass Hierarchy, CP Violation, and Non-Standard Interactions}}

\author[1]{Hridya Harish Nambiar}
\author[1]{Bipin Singh Koranga%
\thanks{Corresponding author. E-mail: bskoranga@kmc.du.ac.in}}
\affil[1]{Department of Physics, Kirori Mal College,
University of Delhi, Delhi~110007, India}
\date{\today}

\begin{document}
\maketitle

\begin{abstract}
We investigate global tripartite quantum entanglement in
three-flavor neutrino oscillations as a tool for probing the
neutrino mass hierarchy and CP violation. Using the linear
entropy formalism, we compute the global entanglement
$\Eg$ for an initial electron neutrino state as a function
of $L/E$, comparing Normal Ordering (NO) and Inverted
Ordering (IO) across CP phases
$\dcp \in \{0^\circ, 90^\circ, 120^\circ, 180^\circ\}$,
in vacuum and in constant-density matter
($\rho = 2.8$~g/cm$^3$, $L = 1300$~km).
We define the hierarchy sensitivity diagnostic
$\DS = \Eg^{\NO} - \Eg^{\IO}$ and show that MSW matter effects
amplify $|\DS|_{\rm max}$ by roughly a factor of two relative
to vacuum, with peak sensitivity at $L/E \approx 655$~km/GeV
($E \approx 2$~GeV at the DUNE baseline).
For antineutrinos the diagnostic is near-perfectly antisymmetric
to the neutrino case at the MSW resonance, with deviations
directly encoding $\dcp$. The combination
$\DSp = \DSn + \DSnb$ and $\DSm = \DSn - \DSnb$
separates the matter hierarchy signal from the CP asymmetry
signal in the tripartite entanglement.
For non-standard interactions (NSI) parameterized by $\varepsilon_{ee}$,
we find $|\DS|_{\rm max} \approx 0.113 + 0.105\,\varepsilon_{ee}$,
a linear and CP-phase-independent relation. The optimal $L/E$
for hierarchy discrimination is stable at $\approx 655$~km/GeV
for all $|\varepsilon_{ee}| \leq 0.2$, providing a robust,
NSI-independent energy recommendation for DUNE.

\end{abstract}

\section{Introduction}
\label{sec:intro}

The question of how quantum information is encoded in
traveling particles has driven a productive intersection between
quantum information theory and neutrino physics over the past
two decades~\cite{Blasone2008,Blasone2009,Jha2021,Abdolalizadeh2024}. Neutrinos are
particularly well-suited for such studies. They travel
macroscopic distances without losing quantum coherence,
interact almost not at all with the environment,
and their flavor oscillations are a pure consequence of quantum
mechanical superposition of mass eigenstates.

The three-flavor neutrino system can be modeled as a three-qubit
system, one qubit per flavor mode, making it a natural setting for
multipartite entanglement studies. Previous work has examined
bipartite entanglement using concurrence and von Neumann
entropy~\cite{Blasone2008,Blasone2014a,Aydiner2023,Koranga2025},
and tripartite entanglement via complementarity
relations~\cite{Bittencourt2022,Bittencourt2023}.
However, none of these studies have simultaneously addressed
CP violation, mass hierarchy discrimination, matter effects,
antineutrinos, and non-standard interactions (NSI)
within a single tripartite entanglement framework.

We close this gap. Using the global linear entropy as our
entanglement measure, we introduce the hierarchy sensitivity
diagnostic $\DS$ and study how it responds to each of the
above effects. We also propose its neutrino-antineutrino
decomposition as a tool to independently isolate the
CP-asymmetry signal and the matter-hierarchy signal.
All results are benchmarked against the DUNE
experiment~\cite{DUNE2016} at $L = 1300$~km.

\section{Theoretical Framework}
\label{sec:theory}

\subsection{Oscillation Probabilities}

Flavor eigenstates $(\nue,\numu,\nutau)$ are related to mass
eigenstates $(\nu_1,\nu_2,\nu_3)$ via the PMNS
matrix~\cite{Pontecorvo1958,Maki1962}:
\begin{equation}
U =
\begin{pmatrix}
c_{12}c_{13} & s_{12}c_{13} & s_{13}e^{-i\dcp} \\
-s_{12}c_{23} - c_{12}s_{23}s_{13}e^{i\dcp} &
 c_{12}c_{23} - s_{12}s_{23}s_{13}e^{i\dcp} & s_{23}c_{13} \\
 s_{12}s_{23} - c_{12}c_{23}s_{13}e^{i\dcp} &
-c_{12}s_{23} - s_{12}c_{23}s_{13}e^{i\dcp} & c_{23}c_{13}
\end{pmatrix},
\end{equation}
where $c_{ij}=\cos\theta_{ij}$, $s_{ij}=\sin\theta_{ij}$.
The vacuum transition probability from $\nue$ to $\nu_\beta$ is
\begin{align}
P(\nue\to\nu_\beta) &= \delta_{e\beta}
-4\sum_{i>j}\mathrm{Re}\bigl[U_{ei}U^*_{\beta i}
U^*_{ej}U_{\beta j}\bigr]
\sin^2\!\!\left(\frac{1.27\,\dm{ij}\,L}{E}\right) \notag \\
&\quad + 2\sum_{i>j}\mathrm{Im}\bigl[U_{ei}U^*_{\beta i}
U^*_{ej}U_{\beta j}\bigr]
\sin\!\!\left(\frac{2.54\,\dm{ij}\,L}{E}\right),
\label{eq:Pvac}
\end{align}
with $\Delta m_{ij}^2$~(eV$^2$), $L$~(km), $E$~(GeV)~\cite{Bilenky1978}.
The second term is the CP-odd contribution.
It vanishes for $\dcp = 0^\circ$ or $180^\circ$
and is largest at $\dcp = 90^\circ$.

\subsection{Matter Hamiltonian and NSI}

Propagation through matter adds a charged-current potential
to the Hamiltonian in the flavor basis
(MSW effect~\cite{Wolfenstein1978,Mikheyev1985}):
\begin{equation}
H_{\rm matter} = H_{\rm vac}
+ \mathrm{diag}\bigl(A(1+\varepsilon_{ee}),\ 0,\ 0\bigr),
\label{eq:Hmatter}
\end{equation}
where $H_{\rm vac} = (2E)^{-1}U\,\mathrm{diag}(0,\dm{21},\dm{31})U^\dagger$
and the MSW potential is
\begin{equation}
A\ (\mathrm{eV}^2) = 7.63\times10^{-5}
\times\rho\ (\mathrm{g/cm}^3)\times Y_e\times E\ (\mathrm{GeV}).
\label{eq:A}
\end{equation}
We use $\rho = 2.8$~g/cm$^3$ and $Y_e = 0.5$ appropriate
for the DUNE Earth-crust baseline.
The parameter $\varepsilon_{ee}$ in Eq.~\eqref{eq:Hmatter}
parameterizes diagonal NSI between electron neutrinos and
matter~\cite{Ohlsson2013}, with current bounds
$|\varepsilon_{ee}|\lesssim 0.3$--$0.5$~\cite{Esteban2018}.
Setting $\varepsilon_{ee}=0$ recovers standard MSW.

Since $A\propto E$, the matter Hamiltonian is energy-dependent
and there is no closed analytic probability formula in the
three-flavor case with $\dcp\neq 0$~\cite{Freund2001}.
At each $L/E$ point we compute $E = L/(L/E)$ with $L=1300$~km,
numerically diagonalize $H_{\rm matter}$, and obtain the
full evolution operator $\mathcal{U}(L) = Ve^{-i\Lambda L}V^\dagger$.
Transition probabilities are then $P_{\alpha\beta}
= |[\mathcal{U}]_{\beta\alpha}|^2$.
For antineutrinos, $A\to -A$ and $U\to U^*$
(equivalent to $\dcp\to-\dcp$),
so whichever ordering is resonance-enhanced for neutrinos
is suppressed for antineutrinos, and vice versa.

\subsection{Entanglement Measure and Hierarchy Diagnostic}

We treat the three flavor modes as qubits.
The initial state $|\nue\rangle = |100\rangle$ evolves into
a superposition of all three flavor states as the neutrino propagates.
The linear entropy of flavor qubit $\alpha$ is
$S_\alpha = 2P_{e\alpha}(1-P_{e\alpha})$,
ranging from 0 (no entanglement) to $1/2$ (maximum entanglement).
The \textit{global tripartite entanglement}
is~\cite{Bittencourt2022}:
\begin{equation}
\Eg = \frac{1}{3}\Bigl[
2P_{ee}(1-P_{ee})
+2P_{e\mu}(1-P_{e\mu})
+2P_{e\tau}(1-P_{e\tau})\Bigr].
\label{eq:Eg}
\end{equation}
This is symmetric over all three flavor subsystems and vanishes
at $L=0$. It is maximized when $P_{e\alpha}=1/3$ for all $\alpha$
(maximum flavor-democratic mixing).

To quantify how well the two mass orderings can be told apart
through entanglement alone, we define the
\textit{hierarchy sensitivity diagnostic}:
\begin{equation}
\DS(L/E) = \Eg^{\NO}(L/E) - \Eg^{\IO}(L/E).
\label{eq:DS}
\end{equation}
When $\DS>0$, NO produces more entanglement at that $L/E$.
When $\DS<0$, IO does. Zeros of $\DS$ are blind spots.
The peak value of $|\DS|$ indicates the best $L/E$
at which a measurement of $\Eg$ can distinguish the two orderings.

For the combined neutrino-antineutrino system, we define:
\begin{equation}
\DSp = \DSn + \DSnb, \qquad \DSm = \DSn - \DSnb.
\label{eq:decomp}
\end{equation}
Since the matter effect is nearly antisymmetric between
neutrinos and antineutrinos,
$\DSp$ largely cancels the matter-driven asymmetry
and exposes the CP signal,
while $\DSm$ largely cancels the CP contribution
and exposes the pure matter hierarchy signal.

\subsection{Oscillation Parameters}
\label{sec:params}

We use the following parameters, consistent with the
NuFIT global fit~\cite{Esteban2020} and DUNE's design values~\cite{DUNE2016}:

\medskip
\begin{center}
\begin{tabular}{ll}
\toprule
Parameter & Value \\
\midrule
$\theta_{12}$ & $33.41^\circ$ \\
$\theta_{13}$ & $8.54^\circ$ \\
$\theta_{23}$ & $49.1^\circ$ \\
$\Delta m_{21}^2$ & $7.37\times10^{-5}$~eV$^2$ \\
$|\dm{31}|$ & $2.56\times10^{-3}$~eV$^2$
(NO: $+$, IO: $-$) \\
$\rho$ & $2.8$~g/cm$^3$ \\
$L$ & $1300$~km (DUNE) \\
$\dcp$ & $0^\circ,\ 90^\circ,\ 120^\circ,\ 180^\circ$ \\
\bottomrule
\end{tabular}
\end{center}
\medskip

The $L/E$ range is $1$--$35000$~km/GeV sampled at 1500 points.
NSI scans cover $\varepsilon_{ee}\in[-0.2,+0.2]$
and $\pm0.5$ as a large-NSI test.

\section{Vacuum Results}
\label{sec:vacuum}

\begin{figure}[H]
\centering
\begin{subfigure}[b]{0.48\textwidth}
    \includegraphics[width=\textwidth]{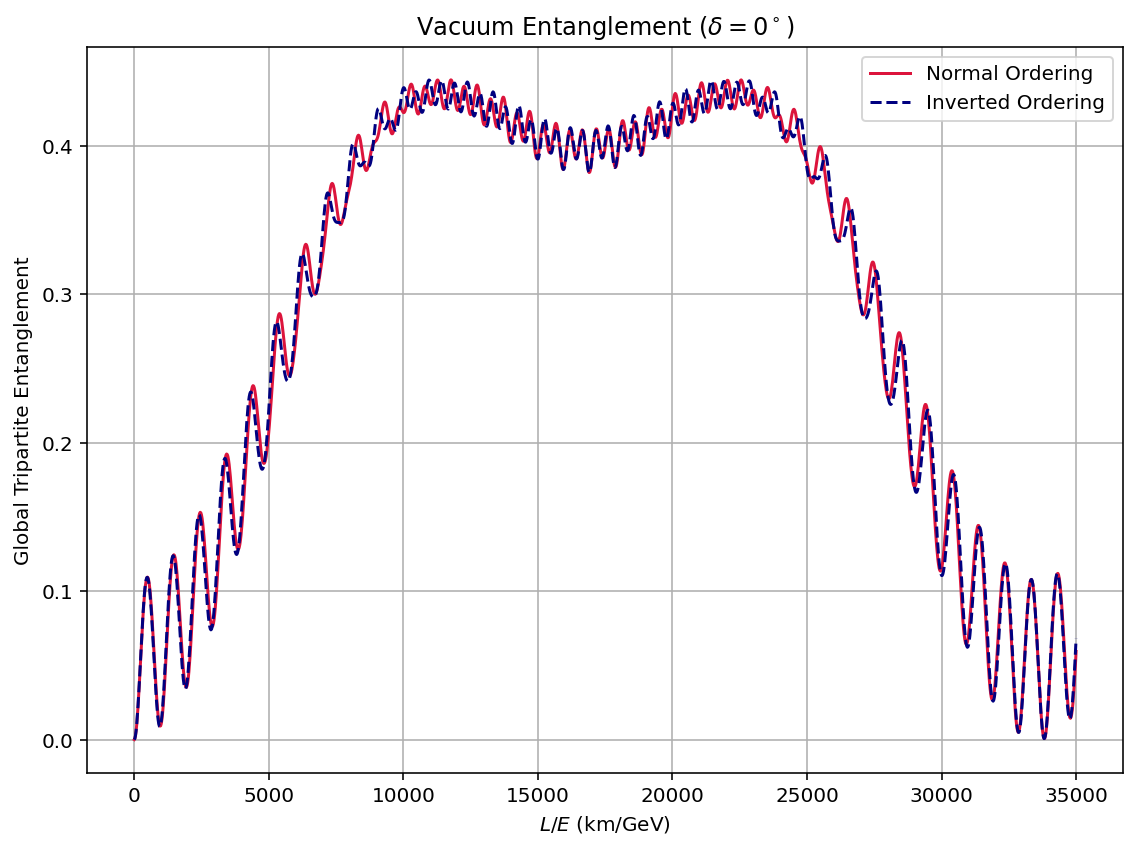}
    \caption{$\dcp=0^\circ$: NO and IO nearly identical.}
\end{subfigure}
\hfill
\begin{subfigure}[b]{0.48\textwidth}
    \includegraphics[width=\textwidth]{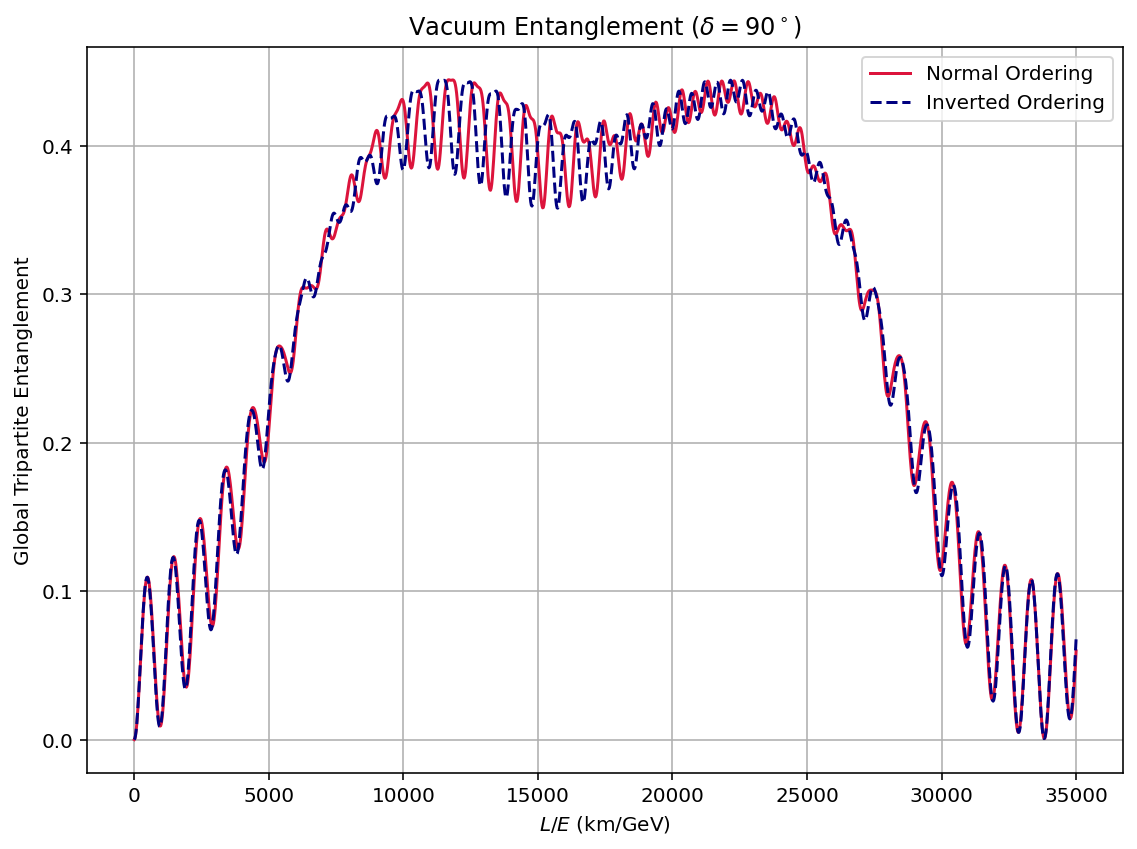}
    \caption{$\dcp=90^\circ$: clear NO/IO splitting.}
\end{subfigure}
\caption{Global tripartite entanglement $\Eg$ vs $L/E$
in vacuum. NO: solid red; IO: dashed blue.
The CP-odd interference drives the NO/IO separation at $90^\circ$.}
\label{fig:vac_01}
\end{figure}

\begin{figure}[H]
\centering
\begin{subfigure}[b]{0.48\textwidth}
    \includegraphics[width=\textwidth]{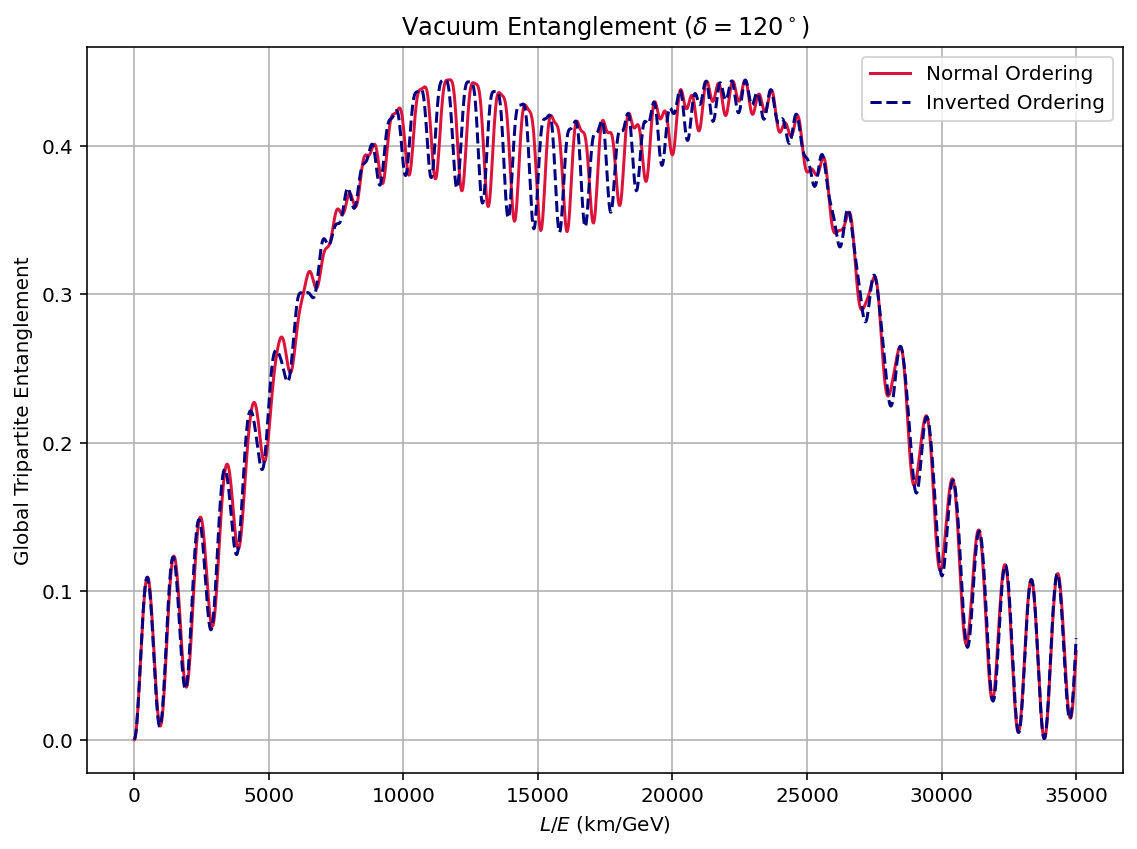}
    \caption{$\dcp=120^\circ$: reduced but visible splitting.}
\end{subfigure}
\hfill
\begin{subfigure}[b]{0.48\textwidth}
    \includegraphics[width=\textwidth]{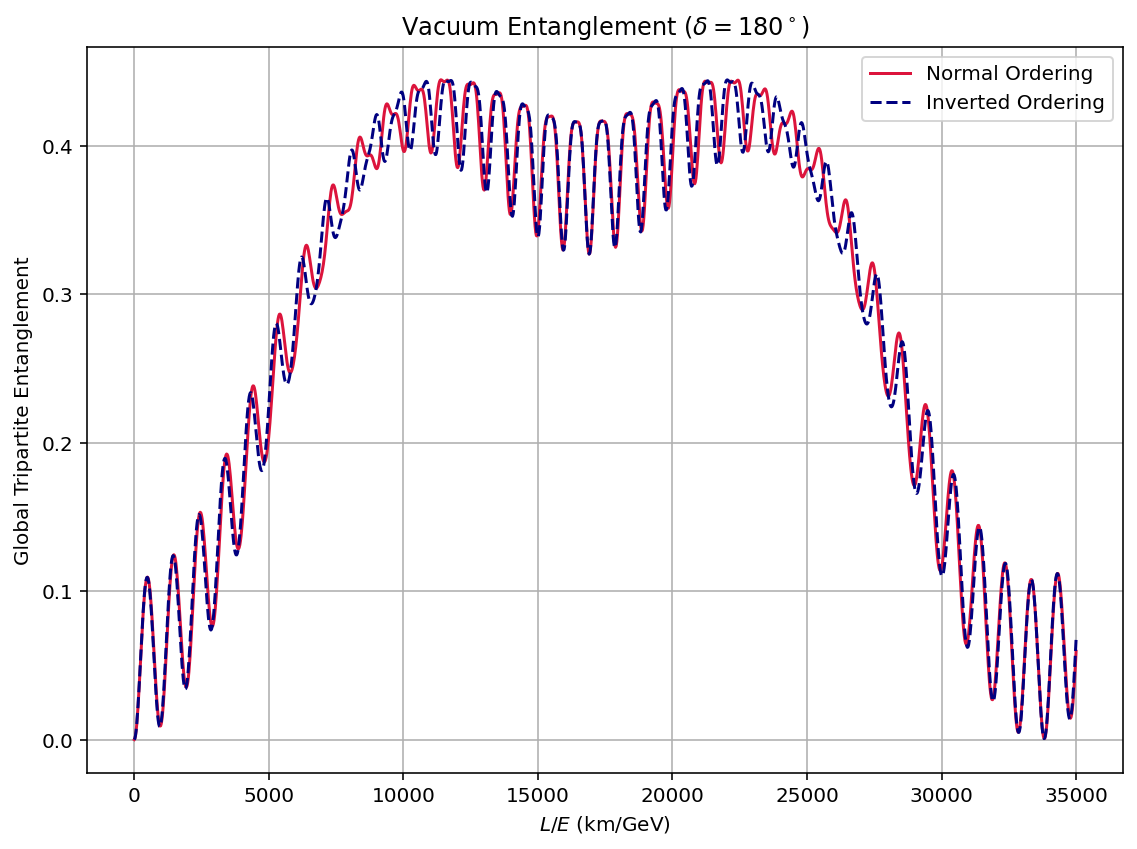}
    \caption{$\dcp=180^\circ$: NO/IO nearly degenerate again.}
\end{subfigure}
\caption{As Figure~\ref{fig:vac_01} for $\dcp=120^\circ$
and $180^\circ$. The splitting follows $|\sin\dcp|$,
vanishing at CP-conserving phases.}
\label{fig:vac_23}
\end{figure}

Figure~\ref{fig:vac_01} shows $\Eg$ versus $L/E$
for $\dcp=0^\circ$ and $90^\circ$ in vacuum;
Figure~\ref{fig:vac_23} shows $120^\circ$ and $180^\circ$.

At $L/E=0$ the neutrino is a pure $|\nue\rangle$ state
and $\Eg=0$. As oscillations build up, $\Eg$ rises and
settles into a broad plateau of $\approx0.43$--$0.45$
spanning $L/E\sim8000$--$25000$~km/GeV,
the regime where flavor probability is spread most evenly
across all three modes. The plateau envelope is shaped by
the slow $\Delta m_{21}^2$driven modulation;
the rapid fine oscillations riding on top of it
are the faster $\Delta m_{31}^2$ interference.
At large $L/E$ the slow envelope brings $\Eg$ back toward zero
as the neutrino returns predominantly to the $\nue$ state.

The effect of the CP phase is clearly visible in the comparison.
At $\dcp=0^\circ$ the CP-odd term in Eq.~\eqref{eq:Pvac} is zero
and NO and IO differ only through the sign of $\dm{31}$,
producing only a tiny phase shift in the fine oscillation structure.
The two ordering curves are almost on top of each other.
At $\dcp=90^\circ$, where the CP-odd term reaches its maximum,
the NO and IO curves develop a clearly visible separation
in the intermediate $L/E$ range.
The NO curve reaches its broad peak at a lower $L/E$
than the IO curve, and the two maintain an offset
throughout the plateau.
This separation arises because the CP-odd term,
proportional to $\mathrm{Im}[\ldots]\sin(2.54\,\dm{ij}\,L/E)$,
redistributes probability between $P_{e\mu}$ and $P_{e\tau}$
in opposite directions for NO and IO
(since $\dm{31}$ appears in the argument with opposite signs).
This asymmetric redistribution propagates into $\Eg$
through Eq.~\eqref{eq:Eg} and produces the visible splitting.
At $\dcp=120^\circ$ the same splitting pattern is present
but with reduced amplitude since $\sin120^\circ<1$.
At $\dcp=180^\circ$ the CP-odd term vanishes again
($\sin180^\circ=0$) and the curves nearly merge,
as at $\dcp=0^\circ$.

\section{Matter Effects: Neutrino Case}
\label{sec:matter}

\subsection{Matter vs Vacuum Comparison}

\begin{figure}[H]
\centering
\begin{subfigure}[b]{0.48\textwidth}
    \includegraphics[width=\textwidth]{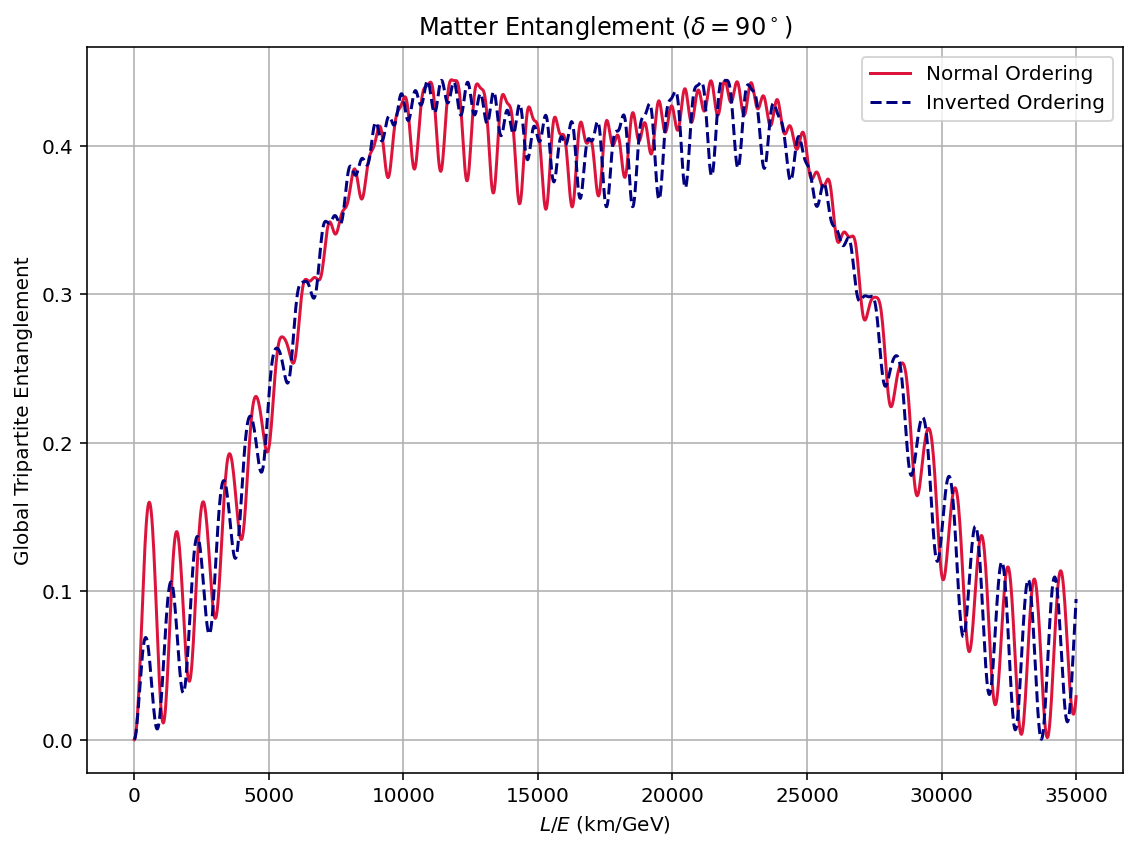}
    \caption{Matter: NO (red) vs IO (dashed blue),
    $\dcp=90^\circ$.}
\end{subfigure}
\hfill
\begin{subfigure}[b]{0.48\textwidth}
    \includegraphics[width=\textwidth]{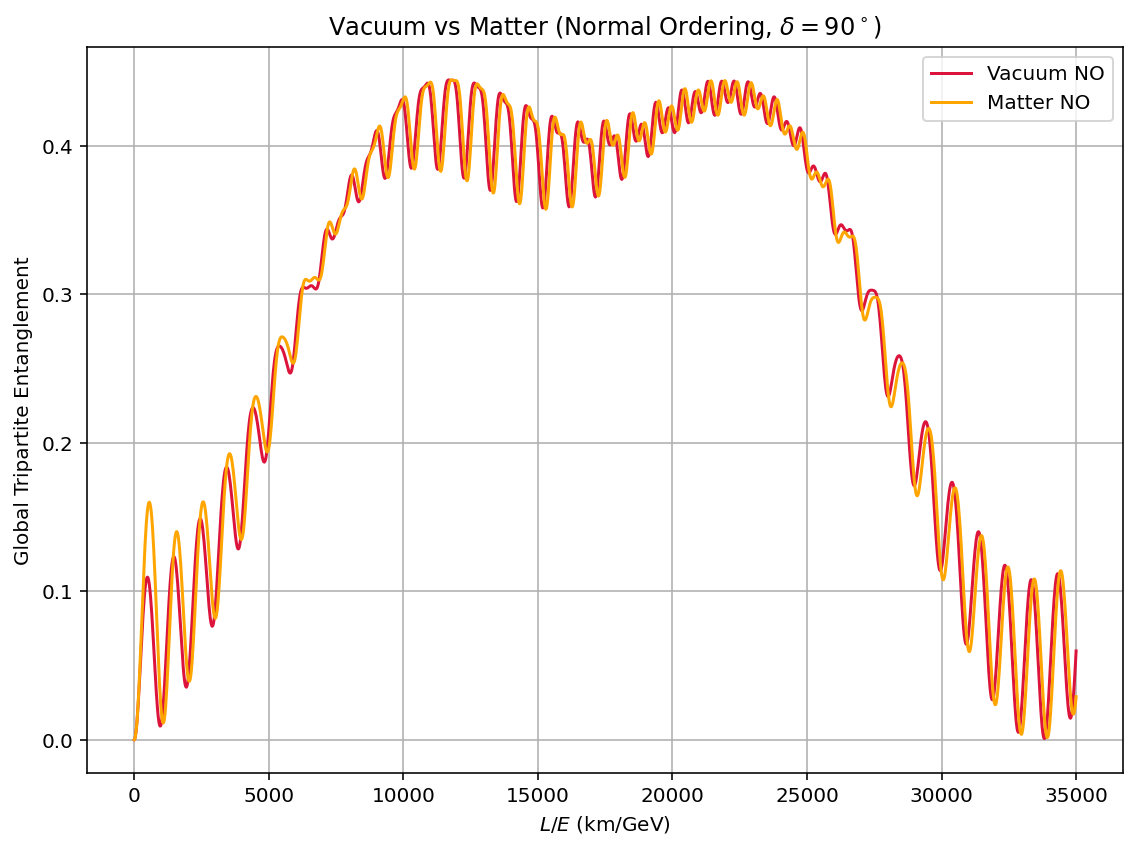}
    \caption{Vacuum (red) vs matter (orange) for NO.}
\end{subfigure}
\caption{Matter entanglement at $\dcp=90^\circ$.
The MSW resonance produces a pronounced elevated peak for NO
at low $L/E$ in matter, absent in vacuum and absent for IO.}
\label{fig:matter_main}
\end{figure}

\begin{figure}[H]
\centering
\includegraphics[width=0.52\textwidth]{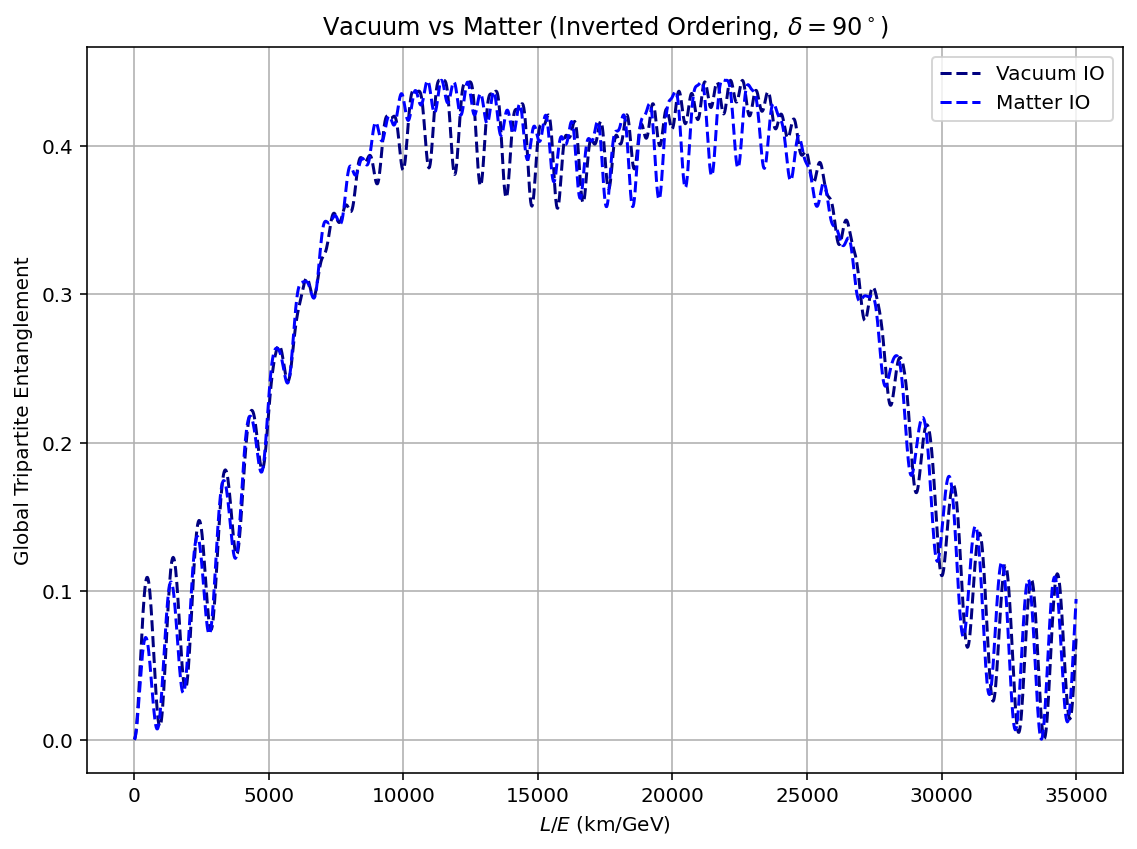}
\caption{Vacuum vs matter for IO at $\dcp=90^\circ$.
The two curves are nearly identical: no MSW resonance
occurs for IO with neutrinos in ordinary matter.}
\label{fig:matter_IO}
\end{figure}

Figure~\ref{fig:matter_main} shows $\Eg$ in matter
for $\dcp=90^\circ$ (left), and the direct vacuum-vs-matter
comparison for NO (right).
Figure~\ref{fig:matter_IO} shows the same comparison for IO. While the CP phase dependence is explored in detail in vacuum, 
the matter analysis is primarily presented for $\delta = 90^\circ$, 
as the CP effecrs will be maximasied for this case. The other values of $\delta$ does not alter the behaviour, and hence are not shown here.

One of the results we have obtained is the elevated first peak
of the NO curve in matter at low $L/E$.
This is the MSW resonance.
The resonance condition for the dominant atmospheric-scale
oscillation is approximately
$A \approx \dm{31}\cos2\theta_{13}/(2E)$,
satisfied for neutrinos in ordinary matter only when
$\dm{31}>0$, i.e.\ only for NO.
Near the resonance energy ($E_{\rm res}\approx10$--$15$~GeV,
corresponding to low $L/E$ at DUNE's 1300~km baseline),
the effective mixing angle $\theta_{13}^{\rm eff}$ is
resonantly driven toward $45^\circ$,
greatly amplifying the $\nue\to\nutau$ oscillation amplitude.
The result is that the NO entanglement receives a strong
boost at low $L/E$ in matter that has no equivalent for IO,
this happens because IO cannot satify the resonance condition for neutrinos.

Looking at the IO comparison in Figure~\ref{fig:matter_IO},
the vacuum and matter curves are almost indistinguishable.
The small differences at intermediate $L/E$ come from
off-resonance matter corrections, which are much weaker
than the resonance enhancement.
This asymmetric response --- large matter effect for NO,
negligible for IO --- is precisely why matter is so valuable
for hierarchy determination, and it is what our
diagnostic $\DS$ is designed to capture.

\subsection{The Hierarchy Sensitivity Diagnostic $\DS$}

\begin{figure}[H]
\centering
\begin{subfigure}[b]{0.48\textwidth}
    \includegraphics[width=\textwidth]{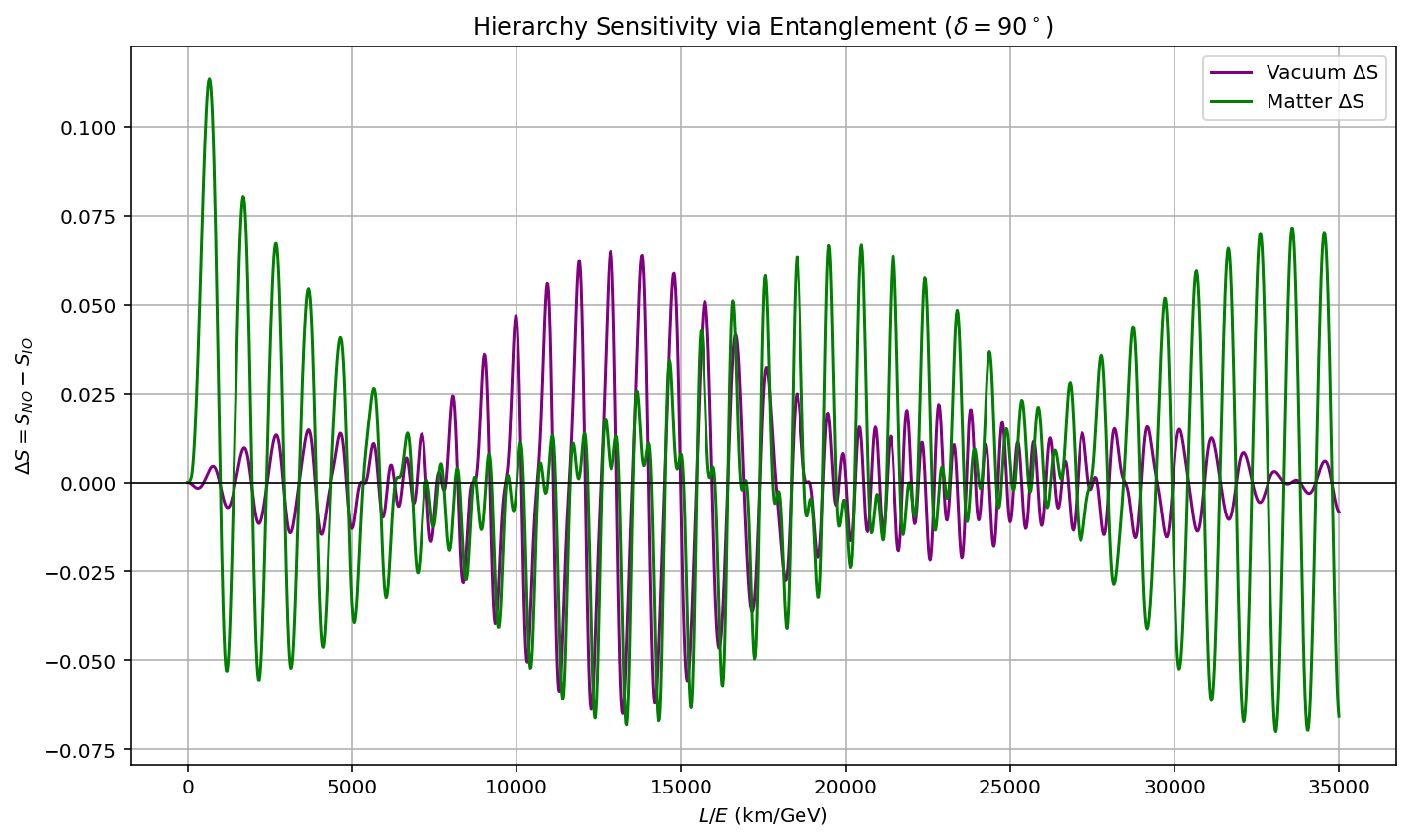}
    \caption{Vacuum (purple) vs matter (green), $\dcp=90^\circ$.}
\end{subfigure}
\hfill

\caption{The hierarchy sensitivity diagnostic $\DS$.
Left: matter doubles the peak sensitivity relative to vacuum
at $L/E\approx655$~km/GeV ($E\approx2$~GeV at DUNE).}
\label{fig:deltaS_pair}
\end{figure}

Figure~\ref{fig:deltaS_pair} shows the central result of
this work: the diagnostic $\DS$ for vacuum and matter
at $\dcp=90^\circ$ (left),
and its CP-phase dependence in matter (right).

\textbf{Matter amplifies hierarchy sensitivity by $\sim2\times$.}
The matter $\DS$ (green) peaks at $|\DS|_{\rm max}\approx0.113$
near $L/E\approx655$~km/GeV.
The vacuum $\DS$ (purple) peaks at $\approx0.06$
in the intermediate range.
The roughly twofold amplification is a direct consequence
of the MSW resonance lifting $\Eg^{\NO}$ while
$\Eg^{\IO}$ stays nearly unchanged,
causing the difference to approximately double at this $L/E$.

\textbf{The optimal point aligns with DUNE.}
The peak sensitivity at $L/E\approx655$~km/GeV corresponds to
a neutrino energy of
\begin{equation}
E_{\rm opt} = \frac{1300\ \rm km}{655\ \rm km/GeV}
\approx 2\ \mathrm{GeV}
\end{equation}
at the DUNE baseline.
DUNE's beam peaks at $\approx2.5$~GeV and carries
significant flux at 2~GeV, making this
an experimentally relevant and accessible energy.

\textbf{The oscillatory structure of $\DS$ is a quantum beating pattern.}
$\DS$ oscillates and changes sign throughout the $L/E$ range.
The oscillatory structure reflects interference between NO and IO frequencies.
The NO and IO entanglement curves oscillate with slightly
different effective frequencies because the sign of $\dm{31}$
shifts the oscillation phase differently in each case.
Their point-by-point difference $\DS$ therefore exhibits
a modulated, oscillating envelope.
Matter sharpens this beating by making the effective NO and IO
oscillation frequencies more distinct near the resonance,
which is why the matter $\DS$ has larger amplitude swings
than the vacuum $\DS$.
The zero-crossings of $\DS$ mark the blind spots where
the two orderings produce identical entanglement
and measurement cannot distinguish them.
The peaks are the sweet spots.

\textbf{CP-phase dependence of $\DS$ in matter.}
The right panel of Figure~\ref{fig:deltaS_pair} shows $\DS$
for $\dcp=0^\circ$, $90^\circ$, and $180^\circ$ in matter.
All three CP phases yield a similar large first peak near
$L/E\approx655$~km/GeV,
confirming that this peak is driven primarily by the
MSW resonance rather than by CP violation --- it is
present regardless of the value of $\dcp$.
In the intermediate $L/E$ range the three CP phases diverge.
Interestingly, $\dcp=180^\circ$ (green) produces the largest
oscillation amplitude in this range.
This is because the CP-even term $\propto\cos\dcp$ changes sign
at $180^\circ$, and its interaction with the matter potential
produces a nontrivial enhancement compared to $0^\circ$ or $90^\circ$
at these intermediate energies.
This is a genuine matter-CP interference effect
and is worth highlighting: it means the choice of $\dcp$
matters most not at the MSW peak but at the lower-energy
intermediate range.

\section{Antineutrino Results and the $\DSp$/$\DSm$ Decomposition}
\label{sec:antineutrino}

\begin{figure}[H]
\centering
\includegraphics[width=0.70\textwidth]{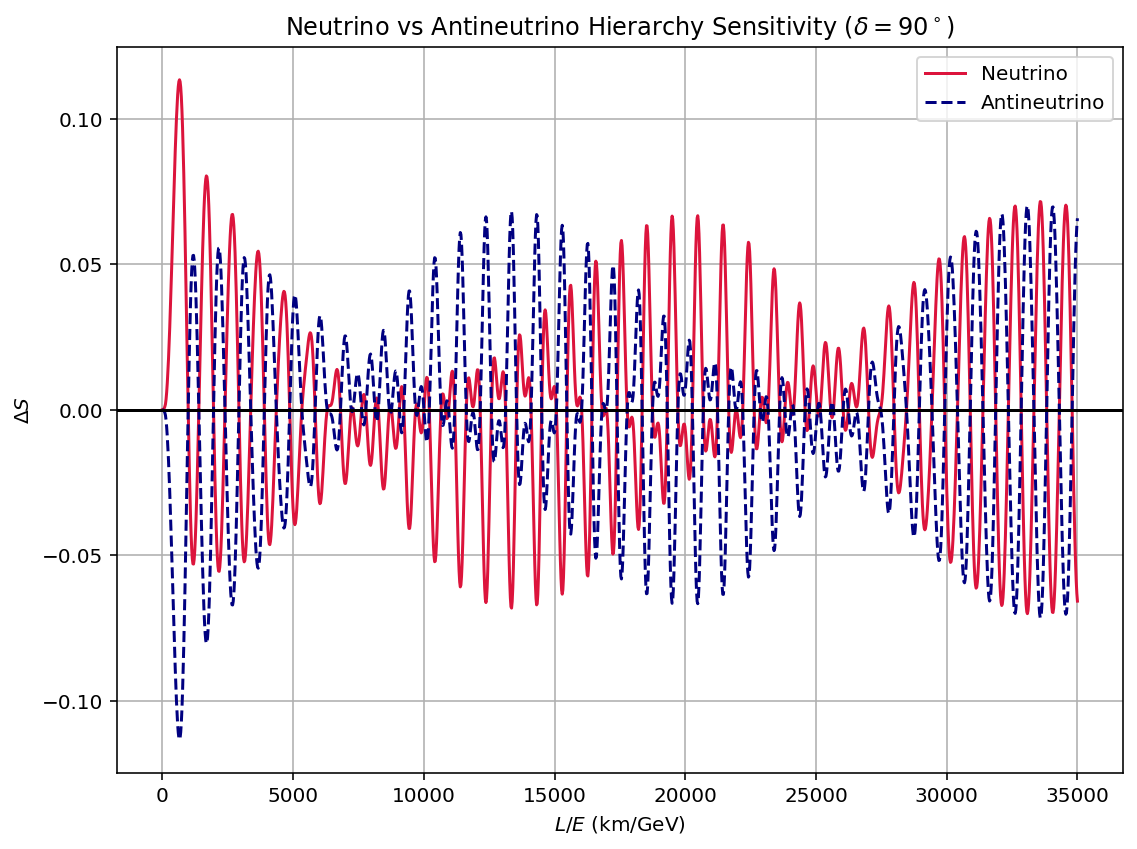}
\caption{$\DS$ for neutrinos (solid red) and antineutrinos
(dashed blue) in matter at $\dcp=90^\circ$.
The near-perfect antisymmetry at low $L/E$ is the MSW
resonance flip. Deviations from exact antisymmetry encode
the CP-violating phase.}
\label{fig:antinu}
\end{figure}

\begin{figure}[H]
\centering
\includegraphics[width=0.70\textwidth]{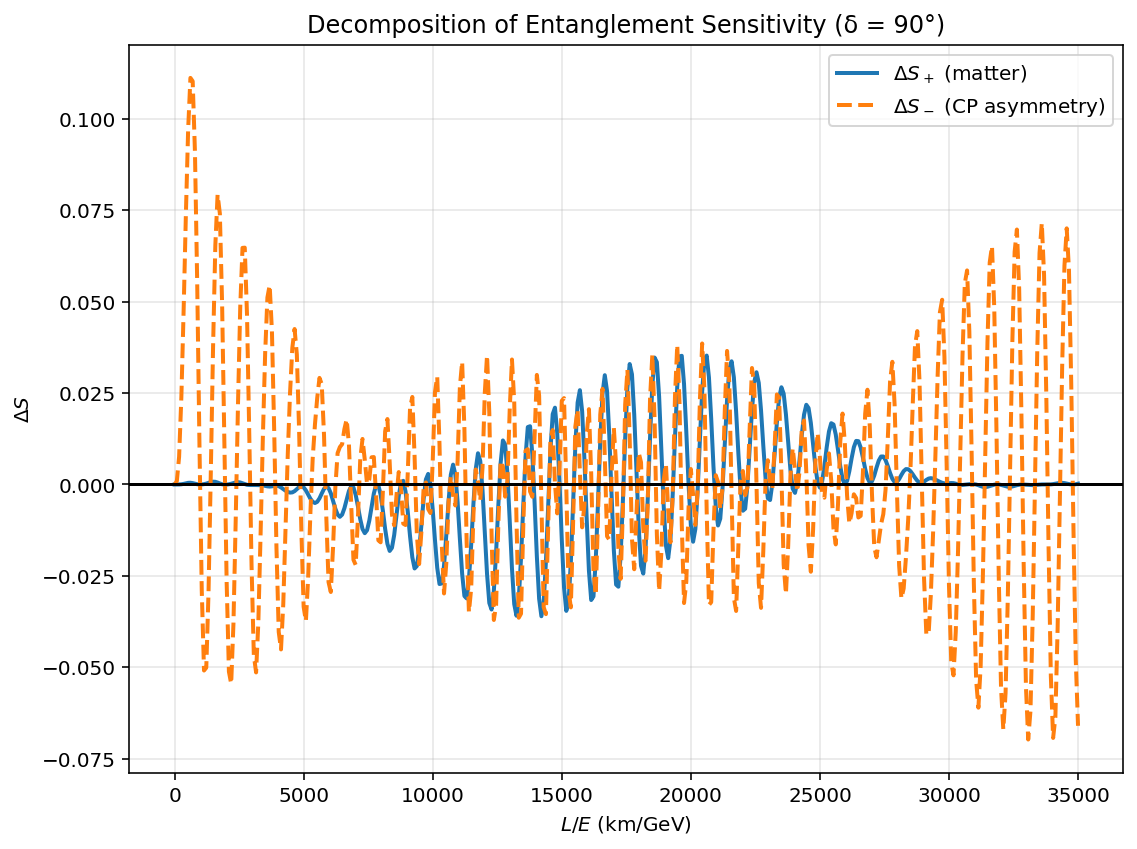}
\caption{Decomposition of the hierarchy sensitivity:
$\DSp = \DSn + \DSnb$ (blue solid, matter signal)
and $\DSm = \DSn - \DSnb$ (orange dashed, CP asymmetry),
at $\dcp=90^\circ$.
The two quantities peak at different $L/E$ ranges,
providing complementary probes.}
\label{fig:decomp}
\end{figure}

Figure~\ref{fig:antinu} shows $\DS$ for neutrinos (red solid)
and antineutrinos (blue dashed) in matter at $\dcp=90^\circ$,
and Figure~\ref{fig:decomp} shows the $\DSp$/$\DSm$ decomposition.

The antineutrino $\DS$ at low $L/E$ is nearly a
mirror image of the neutrino $\DS$ --- it dips to
$\approx-0.113$ where the neutrino peaks at $\approx+0.113$.
This antisymmetry is the clearest possible signal of
the MSW resonance flip.
For antineutrinos $A\to -A$, so the resonance condition
$A\approx\dm{31}\cos2\theta_{13}$ is now satisfied by IO
(since $\dm{31}<0$ for IO and we need $A<0$).
IO is resonance-enhanced for antineutrinos exactly as NO is
for neutrinos.
The result is $\Eg^{\IO}>\Eg^{\NO}$ for antineutrinos,
so $\DS$ is negative --- the hierarchy ordering of the
entanglement has flipped.
The amplitude is the same because the underlying resonance
physics is the same; only the sign flips.

The anti symmetry is not mathematically exact or accurate.
If it were, the neutrino and antineutrino cases would be
related by a pure $\NO\leftrightarrow\IO$ swap and the
curves would be exactly $\DS_{\bar\nu} = -\DS_\nu$.
The small but visible deviations from this at low $L/E$
are the fingerprint of $\dcp\neq0$.
When $\dcp=0^\circ$ the PMNS matrix is real and the
symmetry $U\to U^* = U$ holds exactly,
giving a perfectly antisymmetric pair.
For $\dcp=90^\circ$, the conjugation $U\to U^*$ effectively
sends $\dcp\to-\dcp$, introducing a CP-driven asymmetry
in the oscillation probabilities that breaks the exact mirror
relation between $\DS_\nu$ and $\DS_{\bar\nu}$.
The magnitude of this breaking is a direct measure of $\dcp$.

At intermediate $L/E$ (roughly 5000--25000~km/GeV) the
neutrino and antineutrino $\DS$ curves oscillate with
different phases and amplitudes,
reflecting the combined effect of the opposite matter potential
and the conjugated CP phase.
Neither curve is a simple rescaling of the other in this range.

The $\DSp$/$\DSm$ decomposition in Figure~\ref{fig:decomp}
extracts these two signals cleanly.

$\DSp = \DSn + \DSnb$ (matter signal, blue) is nearly zero
at low $L/E$.
This is expected: the neutrino matter effect (lifting NO)
and the antineutrino matter effect (lifting IO)
are nearly equal and opposite, so they cancel in the sum.
The fact that $\DSp$ is not exactly zero at low $L/E$
confirms the CP-driven breaking of the exact antisymmetry.
$\DSp$ grows and becomes non-negligible at intermediate $L/E$,
reaching amplitudes of $\approx0.025$--$0.035$.
This is the region where the imperfect cancellation between
the neutrino and antineutrino matter effects becomes visible,
and where $\DSp$ provides information about both the hierarchy
and $\dcp$ simultaneously.

$\DSm = \DSn - \DSnb$ (CP asymmetry, orange dashed)
is dominant throughout. At low $L/E$, since
$\DSnb\approx-\DSn$, we have $\DSm\approx 2\DSn$,
which explains why the orange curve has amplitude
$\approx2\times$ the individual neutrino $\DS$.
$\DSm$ is the quantity most sensitive to the hierarchy
at the MSW resonance peak, and it is large wherever
the neutrino $\DS$ is large.

The key experimental implication is that $\DSm$ and $\DSp$
have complementary $L/E$ dependences.
A DUNE measurement at $E\approx2$~GeV (low $L/E$)
primarily probes the CP asymmetry through $\DSm$.
Measurements at higher $L/E$ (lower energies) provide
access to the pure matter hierarchy signal through $\DSp$.
Running DUNE in both neutrino and antineutrino modes and
forming these combinations would allow the two effects
to be disentangled within the tripartite entanglement framework.

\section{Non-Standard Interactions}
\label{sec:nsi}

\begin{figure}[H]
\centering
\begin{subfigure}[b]{0.48\textwidth}
    \includegraphics[width=\textwidth]{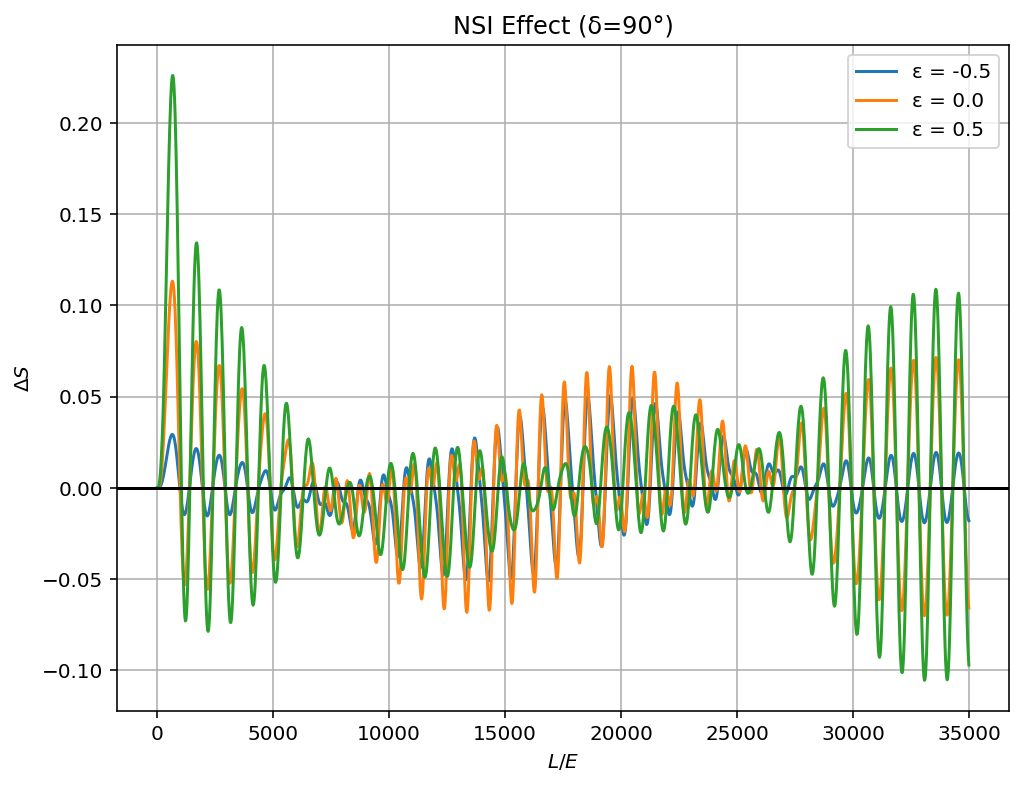}
    \caption{$\DS$ for $\varepsilon_{ee}=-0.5,0.0,+0.5$.}
\end{subfigure}
\hfill
\begin{subfigure}[b]{0.48\textwidth}
    \includegraphics[width=\textwidth]{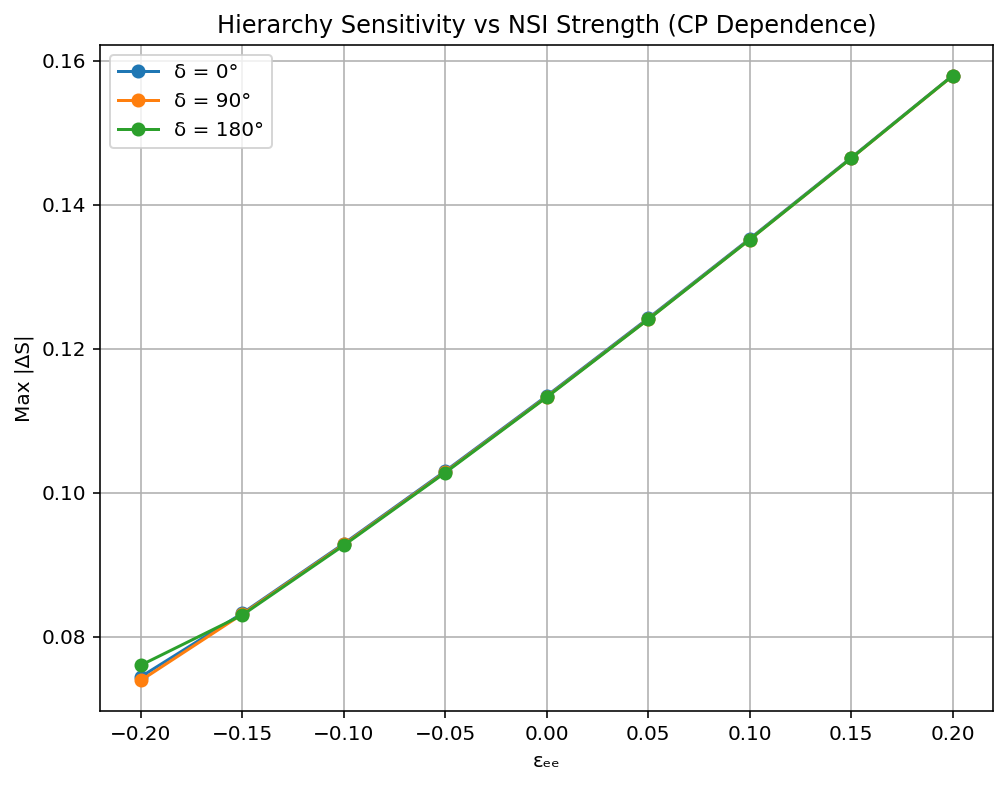}
    \caption{$|\DS|_{\rm max}$ vs $\varepsilon_{ee}$
    for three CP phases.}
\end{subfigure}
\caption{NSI results. Left: positive NSI amplifies the
hierarchy sensitivity while negative NSI suppresses it,
with an asymmetric response from the resonance nonlinearity.
Right: $|\DS|_{\rm max}$ scales linearly with $\varepsilon_{ee}$
following $0.113+0.105\,\varepsilon_{ee}$,
independent of the CP phase.}
\label{fig:nsi_curves}
\end{figure}

\begin{figure}[H]
\centering
\begin{subfigure}[b]{0.48\textwidth}
    \includegraphics[width=\textwidth]{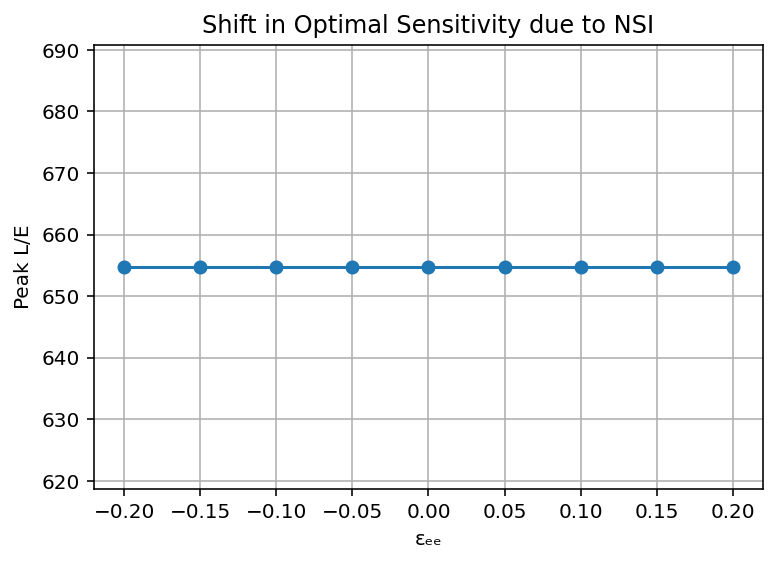}
    \caption{Optimal $L/E$ vs $\varepsilon_{ee}$: completely flat.}
    \label{fig:nsi_loc}
\end{subfigure}
\hfill
\begin{subfigure}[b]{0.48\textwidth}
    \includegraphics[width=\textwidth]{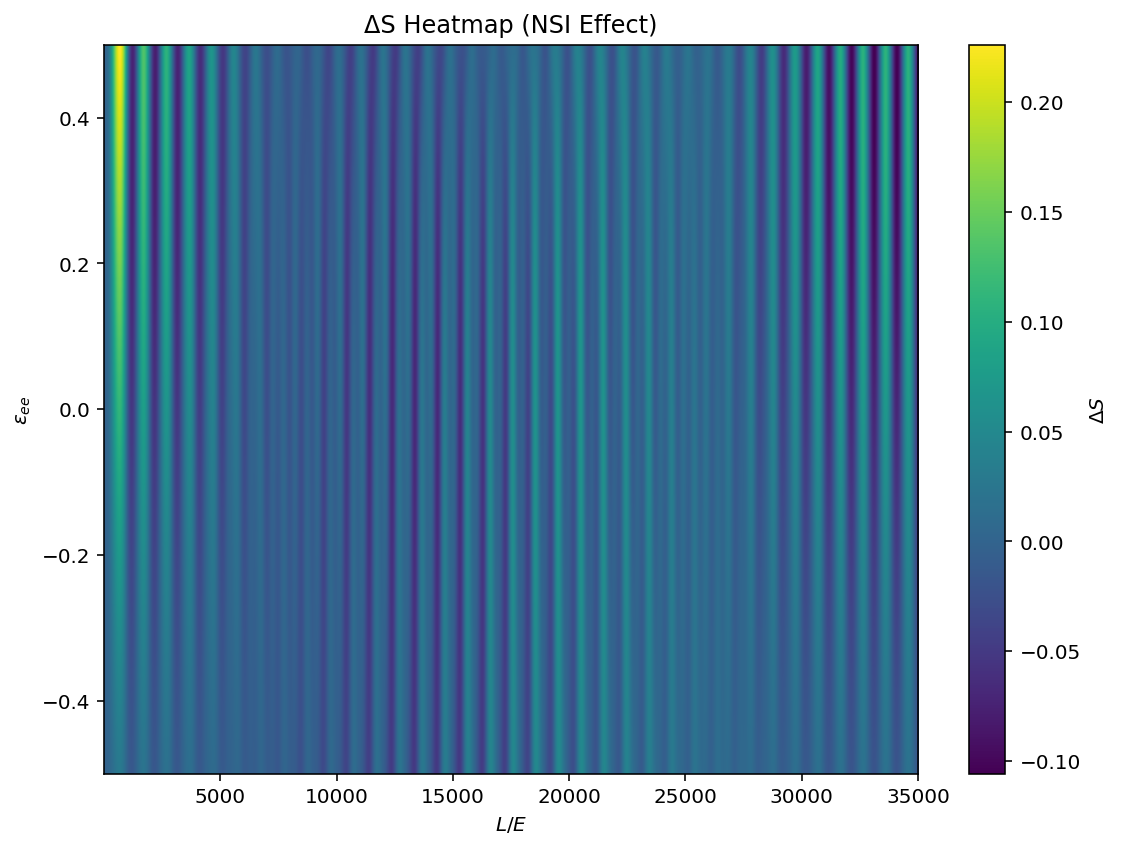}
    \caption{Full $\DS$ heatmap over $\varepsilon_{ee}$ and $L/E$.}
    \label{fig:nsi_heat}
\end{subfigure}
\caption{Left: the optimal $L/E$ for hierarchy discrimination
is pinned at $\approx655$~km/GeV regardless of NSI strength,
providing an NSI-independent energy recommendation for DUNE.
Right: the heatmap shows that NSI scales the amplitude
of $\DS$ uniformly without shifting the $L/E$ structure.}
\end{figure}

Figure~\ref{fig:nsi_curves} shows $\DS$ for
$\varepsilon_{ee}\in\{-0.5,0.0,+0.5\}$ at $\dcp=90^\circ$ (left),
and $|\DS|_{\rm max}$ vs $\varepsilon_{ee}$ for three
CP phases (right).
Figures~\ref{fig:nsi_loc} and~\ref{fig:nsi_heat} show the
optimal $L/E$ stability and the full $\DS$ heatmap.

\textbf{Amplitude scaling with NSI.}
Positive $\varepsilon_{ee}$ enhances the effective matter
potential $V_{ee}=A(1+\varepsilon_{ee})$, deepening the
MSW resonance asymmetry between NO and IO.
At $\varepsilon_{ee}=+0.5$ the first peak of $\DS$
reaches $\approx0.22$ --- nearly double the standard-matter value.
At $\varepsilon_{ee}=-0.5$ the potential is halved;
the resonance enhancement is substantially reduced and the
first peak drops to $\approx0.03$,
rendering the entanglement almost insensitive to the hierarchy
at this $L/E$.
The response is clearly asymmetric: doubling the potential
approximately doubles the peak $\DS$,
while halving it reduces the peak by a factor of $\approx4$.
This asymmetry arises from the nonlinearity of the resonance
condition --- positive NSI sharpens the resonance while
negative NSI broadens and eventually eliminates it.

\textbf{Linear scaling of $|\DS|_{\rm max}$.}
The right panel of Figure~\ref{fig:nsi_curves} shows that
$|\DS|_{\rm max}$ increases linearly with $\varepsilon_{ee}$:
\begin{equation}
|\DS|_{\rm max} \approx 0.113 + 0.105\,\varepsilon_{ee}.
\label{eq:linear}
\end{equation}
The slope is $\approx0.105$ per unit $\varepsilon_{ee}$,
meaning every increase of $\Delta\varepsilon_{ee}=0.1$
boosts the maximum hierarchy sensitivity by approximately $0.0105$.
More remarkably, all three CP phases ($0^\circ$, $90^\circ$, $180^\circ$)
fall on exactly the same line.
The NSI-driven enhancement of $|\DS|_{\rm max}$ is
completely independent of the CP-violating phase.
This is a clean and perhaps surprising result:
while the CP phase determines the detailed shape of $\DS$
over $L/E$ and controls where the intermediate-range
sensitivity is concentrated,
it has no effect on the overall scaling of the
peak sensitivity with $\varepsilon_{ee}$.
Equation~\eqref{eq:linear} is therefore a quantitative,
CP-blind prediction:
if DUNE measures $\varepsilon_{ee}\neq0$ through
dedicated NSI analyses,
the shift in the maximum entanglement-based hierarchy
sensitivity is given by $0.105\,\varepsilon_{ee}$,
regardless of the simultaneously measured value of $\dcp$.

\textbf{Stability of the optimal $L/E$.}
Figure~\ref{fig:nsi_loc} shows the $L/E$ value at which
$|\DS|$ is maximum, as a function of $\varepsilon_{ee}$.
The curve is completely flat at $\approx655$~km/GeV
across the entire scanned NSI range.
NSI does not move the optimal point; it only changes
how strongly the hierarchy is imprinted on the entanglement there.
This means the experimental recommendation for DUNE
--- target $E\approx2$~GeV for maximum hierarchy sensitivity
via tripartite entanglement --- is completely robust
against non-standard interactions within current bounds.

\textbf{The $\DS$ heatmap.}
Figure~\ref{fig:nsi_heat} shows the full two-dimensional
$\DS$ landscape over $\varepsilon_{ee}$ and $L/E$.
The dominant pattern is vertical stripes,
confirming that the oscillation structure in $L/E$ is
preserved across all NSI values.
The bright yellow region at far left is the MSW resonance peak,
which intensifies with increasing $\varepsilon_{ee}$
as expected from the linear scaling in Eq.~\eqref{eq:linear}.
The horizontal invariance of the stripe pattern
is a visual confirmation that NSI acts as a
uniform brightness dial on the existing sensitivity landscape
rather than reshaping it.

\section{Conclusions}
\label{sec:conclusions}

We have studied global tripartite quantum entanglement in
three-flavor neutrino oscillations as a probe of the
mass hierarchy and CP violation,
extending previous work to include the CP-phase dependence,
systematic NO/IO comparison, matter effects,
antineutrinos, and non-standard interactions simultaneously.

Our main findings are:

\begin{enumerate}

\item In vacuum, the global entanglement $\Eg$ rises from zero,
forms a broad plateau at $\approx0.43$--$0.45$,
and falls at large $L/E$.
The NO/IO splitting follows $|\sin\dcp|$,
vanishing at $\dcp=0^\circ$ and $180^\circ$ and
reaching its maximum at $\dcp=90^\circ$.

\item Matter effects (MSW) resonantly amplify the NO entanglement
while leaving IO nearly unchanged.
The hierarchy sensitivity diagnostic $\DS$ peaks at
$|\DS|_{\rm max}\approx0.113$ at $L/E\approx655$~km/GeV in matter,
roughly twice the vacuum value.
This optimal point corresponds to $E\approx2$~GeV at the
DUNE baseline, directly within DUNE's operating energy range.

\item For antineutrinos, $\DS$ is near-perfectly antisymmetric
to the neutrino case at the MSW resonance.
The deviation from exact antisymmetry encodes $\dcp$
and grows with its magnitude.

\item The decomposition $\DSp = \DSn + \DSnb$ and
$\DSm = \DSn - \DSnb$ separates the matter hierarchy
signal from the CP asymmetry signal in the tripartite entanglement,
with complementary $L/E$ sensitivity ranges.

\item NSI modifies $|\DS|_{\rm max}$ linearly:
$|\DS|_{\rm max}\approx0.113+0.105\,\varepsilon_{ee}$,
independently of $\dcp$.
The optimal $L/E$ for hierarchy discrimination is stable
at $\approx655$~km/GeV for all $|\varepsilon_{ee}|\leq0.2$.

\end{enumerate}


\end{document}